\documentclass[preprint2]{aastex}
\usepackage{epsfig}
\usepackage{natbib}
\usepackage{amsmath}
\usepackage{multicol}

\newcommand{\kms}{km~s$^{-1}$}
\newcommand{\Msun}{M_{\odot}}

\newcommand{\kmsMpc}{km~s$^{-1}$~Mpc$^{-1}$}

\DeclareMathAlphabet{\mathpzc}{OT1}{pzc}{m}{it}

\begin{document}
\title{The Cosmic V-Web}

\author{Daniel Pomar\`ede}
\affil{Institut de Recherche sur les Lois Fondamentales de l'Univers, CEA/Saclay, 91191 Gif-sur-Yvette, France}
\and
\author{Yehuda Hoffman}
\affil{Racah Institute of Physics, Hebrew University, Jerusalem, 91904 Israel}
\and
\author{H\'el\`ene M. Courtois}
\affil{Universit\'e Claude Bernard Lyon I, Institut de Physique Nucl\'eaire, Lyon, France}
\and
\author{R. Brent Tully,}
\affil{Institute for Astronomy, University of Hawaii, 2680 Woodlawn Drive,
 Honolulu, HI 96822, USA}

\begin{abstract}
The network of filaments with embedded clusters surrounding voids seen in maps derived from redshift surveys and reproduced in simulations has been referred to as the cosmic web.  A complementary description is provided by considering the shear in the velocity field of galaxies.  The eigenvalues of the shear provide information on whether a region is collapsing in three dimensions, the condition for a knot, expanding in three-dimensions, the condition for a void, or in the intermediate condition of a filament or sheet.  The structures that are quantitatively defined by the eigenvalues can be approximated by iso-contours that provide a visual representation of the cosmic velocity (V) web.  The current application is based on radial peculiar velocities from the {\it Cosmicflows-2} collection of distances.  The three-dimensional velocity field is constructed using the Wiener filter methodology in the linear approximation.  Eigenvalues of the velocity shear are calculated at each point on a grid.  Here, knots and filaments are visualized across a local domain of diameter $\sim 0.1c$.

\smallskip\noindent
Key words: large scale structure of universe --- galaxies: distances and redshifts
\bigskip
\end{abstract}

\smallskip
\section{Introduction}

Structure in the universe was born out of tiny fluctuations in the cauldron of the Big Bang.  Although on large scales cosmic expansion is dominant, locations of higher density attract and locations of lower density evacuate.  If the local density is high enough, attraction can trump cosmic expansion and lead to collapse.  The patterns of density irregularities are complex and at any location there are many influences.  The velocities of galaxies after the cosmic expansion is subtracted out are a reflection of these influences.  With low amplitude density fluctuations, there is a linear relationship between the distribution of matter and the divergence of the velocity field.  This simple relationship invites a way to describe the features of the large-scale structure in the distribution of galaxies: the cosmic velocity or V-web  \citep{2012MNRAS.425.2049H}.  

The shear in the velocity field can be determined at any specified location.  If all three eigenvalues of the shear are positive above some threshold then the location is within the domain of a collapsing knot.  If two of these eigenvalues exceed the threshold then the location is within a filament, collapsing in two dimensions and expanding in the third.  One eigenvalue above the threshold identifies a sheet while values below the threshold on all three axes is the property of a void.  Surfaces can be defined that separate the knot, filament, sheet, and void constituents.  For thresholds with positive values most of the volume of the universe is given over to sheets and voids.  Here we study the cosmography of the nearby universe primarily through representations of the knots and filaments, the regions of highest density.

\section{Context}

Early evidence of sheet-like structure in the distribution of galaxies \citep{1956VA......2.1584D} was given an explanation in seminal discussions of the anisotropic collapse of large scale density perturbations \citep{1970A&A.....5...84Z, 1989ptpa.book..338Z}.  The increasingly clear evidence for filamentary structure from observations \citep{1982ApJ...257..389T, 1986ApJ...302L...1D} and the increasingly sophisticated simulations of clusters connected by filaments and sheets with most of space in voids gave rise to the descriptive terminology of the cosmic web \citep{1996Natur.380..603B}.

Various schemes have been developed to characterize the features of the cosmic web in the distribution of galaxies in redshift space or the distribution of particles in simulations.  Among them are DisPerSE \citep{2011MNRAS.414..350S}, SpineWeb \citep{2010ApJ...723..364A},  Bisous \citep{2014A&A...572A...8T}, and the Multiscale Morphology Filter \citep{2007A&A...474..315A, 2010MNRAS.408.2163A}.  NEXUS is an extension of this latter to incorporate velocity signatures, tides, divergences, and shears \citep{2013MNRAS.429.1286C, 2014MNRAS.441.2923C}.  Others have used the tidal tensor in outlining the web \citep{2007MNRAS.375..489H, 2009MNRAS.396.1815F}.  At higher densities the dynamics become non-linear, reaching shell-crossing and phase mixing  \citep{2012MNRAS.427...61A, 2012PhRvD..85h3005S, 2012ApJ...754..126F}.  This paper focuses on the identification of web features from velocities alone and with an analysis restricted to the linear regime \citep{2012MNRAS.425.2049H, 2012MNRAS.421L.137L, 2015MNRAS.453L.108L}.

\section{Cosmicflows-2 Velocities}

The line-of-sight velocities of galaxies can be readily and accurately obtained from the Doppler-shift of spectral lines.  These velocities are the combination of a cosmic expansion component, usually dominant, and a component referred to as the peculiar velocity that is associated with density irregularities.  The cosmic expansion component obeys the Hubble law, linearly increasing with distance, so if a distance to a galaxy is known then the cosmic expansion velocity is inferred and the residual from the observed velocity is the line-of-sight peculiar velocity.

In the volume of interest of this study there are of order a million galaxies.  Accurate distances are available for only a tiny fraction.  Galaxies with measured distances are tracer test particles that sample the departures from a pure Hubble flow.  The present study is based on the {\it Cosmicflows-2} catalog of ~8000 distances \citep{2013AJ....146...86T}.  {\it Cosmicflows-2} distances are derived from six methodologies: (1) the relation between the pulsation period of Cepheid stars and their luminosity, (2) the constrained luminosity of red giant branch stars at the onset of core Helium burning, (3) the so-called surface brightness fluctuation method where the brightest red giant branch stars in a galaxy are individually unresolved but collectively mottle the appearance of the galaxy in a way that diminishes with distance, (4) the so-called Tully-Fisher method that is based on the correlation between the rotation rate of a spiral galaxy and its luminosity, (5) the related fundamental plane method for elliptical galaxies which links luminosity, central velocity dispersion, and surface brightness, and (6) the almost constant peak brightness of supernova explosions of type I$a$.  

This compendium of distances involves both original measurements made within our collaboration and contributions from the literature, in roughly equal parts.  There was an attempt with the assembly of material from different sources to assure a consistency of scale.  The global value for the Hubble Constant consistent with this data was found to be $H_0 = 74.4$~\kmsMpc\ although the zero point scale is not important for the present discussion.  Most recently, our collaboration has published the expanded {\it Cosmicflows-3} catalog of ~18,000 distances \citep{2016AJ....152...50T} but this new catalog remains to be studied.

The observed velocity of a galaxy, $V_{obs} = cz$, at a distance $d$, can be separated into a component due to the expansion of the universe, $H_0 d$, and a peculiar velocity, $V_{pec}$, the radial component of the departure from cosmic expansion \citep{2014MNRAS.442.1117D}
\begin{equation}
V_{pec} = (\mathpzc{f} V_{obs} - H_0 d)/(1 + H_0 d/c)
\label{eq:vpds}
\end{equation}
where 
\begin{equation}
\mathpzc{f} = 1 + {{1}\over{2}} (1 - q_0)z - {{1}\over{6}} (1 - q_0 -3 q_0^2 + j_0)z^2
\label{eq:f}
\end{equation}
and we assume $q_0 = (\Omega_m - 2\Omega_{\Lambda})/2 = -0.595$ ($\Omega_m=0.27$ and flat topology) and $j_0 = 1$.  Details of the cosmological parameters are not important on the local scale of the current study.

Only the radial components of peculiar velocities are observed, and with substantial uncertainties.  The sky coverage is non-uniform, particularly due to obscuration at low galactic latitudes, but also because of the heterogeneous nature of the distance material and decreased density of coverage with redshift.  A crucial element of our program involves the translation of anisotropic, sparse-sampled and noisy radial peculiar velocities into a map of three-dimensional velocities.

\section{Global Velocity Fields}

A Bayesian construction of the three dimensional velocity field is performed by means of the Wiener filter estimator \citep{1995ApJ...449..446Z, 1999ApJ...520..413Z} applied within the framework of the standard model of cosmology. The model presumes that structure formed from primordial Gaussian fluctuations with a specified power spectrum \citep{2009ApJS..180..330K}.  The linear approximation for the response of peculiar velocities holds reasonably well down to a scale of $\sim 5$ Mpc.
Uncertainties in individual distances and hence $V_{pec}$ are substantial, increasingly so at large distances.   A proper analysis must account for velocity field biases \citep{1995PhR...261..271S}.  Here we use the Bayesian methodology of the Wiener filter on the {\it Cosmicflows-2} collection of distances.  The products are an optimal compromise between the noisy and sparse data and the prior assumptions embodied in the cosmological model and linear theory.


The linear Wiener filter is applied to the sparse and noisy radial velocity measurements to produce three-dimensional velocity fields because of long-range coherence in flow patterns.  Indeed, structures beyond the range of the distance information can be inferred from the tidal component of the velocity field.  Dipole and quadrupole moments of the velocity field carry information on attractors and repellers on very large scales \citep{1986ApJ...307...91L, 2001astro.ph..2190H, 2015MNRAS.449.4494H, 2017NatAs...1E..36H}.  The analysis gives attention to Malmquist biases \citep{1995PhR...261..271S}.  There have been extensive discussions of these issues in several publications \citep{1980lssu.book.....P,2008LNP...740..409V, 2009MNRAS.392..743W, 2016IAUS..308..310D}.

\section{Local Velocity Fields and Velocity Shear}
\label{sec:velocity}

Our interest here is on intermediate and local scales.  We invoke two diagnostic tools.  One of these is decomposition of the global flow fields into local and tidal components \citep{1999ApJ...520..413Z,2001astro.ph..2190H}.  In the linear regime the Poisson and Euler equations prescribe that the divergence of the velocity is proportional to the density.   The Wiener filter analysis gives us a 3D map of linear density.  We can extract the density in a given volume about a selected location and set the density outside this region to the mean.  A local velocity field is associated with the extracted density field.  The vector subtraction of the local velocity from the global one gives the tidal velocity field.\footnote{The term `tidal' used here refers to all orders of the multipole influences of gravity outside the local region.}  The origin and the radius of the extraction procedure are freely chosen, allowing flexibility in investigations of suspected basins of attraction and, with the inversion of gravity, repulsion.

Our second diagnostic tool involves an evaluation of the shear in the velocity field on a grid of specified locations.
\begin{equation}
\Sigma_{\alpha\beta} = - (\partial_{\alpha} V_{\beta} + \partial_{\beta} V_{\alpha})/2H_0
\label{eq:shear}
\end{equation}
where partial derivatives of velocity $V$ are determined along directions $\alpha$ and $\beta$ of the orthogonal supergalactic cartesian axes and a normalization is provided by the average expansion rate of the universe given by $H_0$, the Hubble Constant.  The minus sign is introduced to associate positive eigenvalues with collapse.
 
\begin{figure}[]
\begin{center}                                                                                                                                                                                                                                                                                                                                                                                                                                                                                                                                                                                      
\includegraphics[scale=.55]{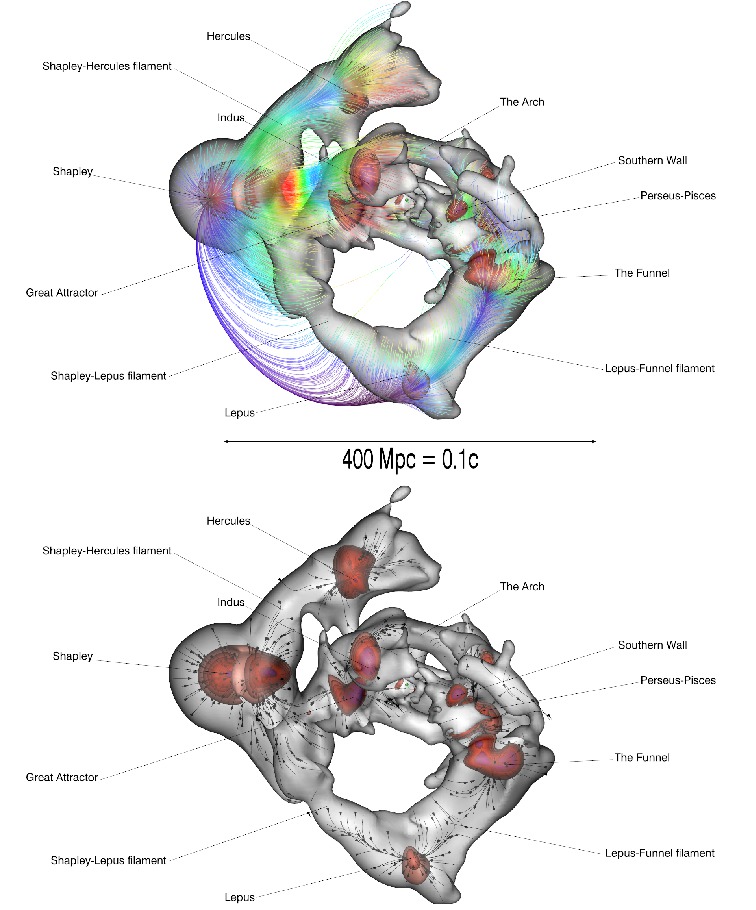}
\caption{The cosmic V-web.  Surfaces of knots, in red, are embedded within surfaces of filaments, in grey.  The top panel illustrates large scale flow patterns from selected seed positions terminating in the Shapley Concentration.  Motions accelerate toward knots (redder tones) and are retarded beyond knots (bluer tones).  The bottom panel shows local flows toward local basins of attraction.  There are frequently divergent points along filaments, with local velocities at adjacent positions in opposite directions.  The scale of the V-web representation approximates $0.1c$ diameter.  The top panel appears at 1:38 in the animated figure and the bottom panel appears at 2:21.}
\label{fig1}
\end{center}                                                                                                                                                                                                                                                                                                                                                                                                                                                                                                                                                                                        
\end{figure}

The eigenvectors of the shear, with associated eigenvalues ordered from most positive to most negative, define the principal axes of the collapse and expansion.  We are most interested here in the regions of highest density; knots with all eigenvalues positive and above a specified threshold, and filaments with two of three eigenvalues above the threshold.  Our visual representations show surfaces that enclose knots and filaments above specified thresholds.  These representations are visual proxies for the velocity-web or, for short, the V-web \citep{2012MNRAS.425.2049H, 2012MNRAS.421L.137L, 2014MNRAS.441.1974L}.  The two panels of Figure~\ref{fig1}\footnote{Each of the still figures in this article are directly or approximately drawn from scenes in the animated figure embedded in the on-line journal article and also available at http://vimeo.com/pomarede/vweb.} illustrate this proxy for the V-web in the 400~Mpc $\sim 0.1c$ diameter local domain, with surfaces of knots in red and surfaces of filaments in grey.  Knots are represented with surfaces at five levels of the smallest eigenvalue $\lambda_3$ (positive at 0.04, 0.05, 0.06, 0.07, and 0.095) while filaments are represented with surfaces at a single choice of the eigenvalue $\lambda_2 = +0.046$.  The streamlines on these plots are alternative descriptions of flows, representing global flows in the top panel and local flows in the bottom panel.

We represent flows with streamlines.  The Hubble expansion has been subtracted out.  A streamline is developed from a selected seed position by constructing a vector of length covered in one time step, then repeating the process at the new position at the end of the vector, step at a time, until reaching a sink or leaving the box.  The line equation of a streamline, $\vec{l}(s)$ where $s$ is the line parameter, is calculated by integrating  $d \vec{l}(s) = \vec{v}(\vec{l}(s)) d s$
(Hoffman et al. 2017).  In the top panel of Fig.~\ref{fig1} with full flows, colors represent velocities that rise as the gravitational potential minima (calculated by solving the Poisson equation) are approached and fall as a minimum is passed.

The bottom panel gives focus to local flows toward local gravitational wells - where local velocities are those induced by the density field within spheres of 6,000~\kms.  It is commonly seen along filaments that there are divergences in local flows, boundaries between flows in opposite directions along the filament.  If the regions of these boundaries are inspected in a mapping of individual galaxies, usually there is an apparent disruption in the continuity of the distribution.  The dimension of the local analysis is arbitrary.  The density field contains a hierarchy of structures \citep{2013MNRAS.428.3409A, 2013MNRAS.429.1286C}. Groups and individual galaxies are regions of collapse within filaments and sheets and even voids.  The immediate regions of collapse around such entities are not captured by the Wiener filter linear analysis.

\begin{figure}[!]
\begin{center}                                                                                                                                                                                                                                                                                                                                                                                                                                                                                                                                                                                      
\includegraphics[scale=.114]{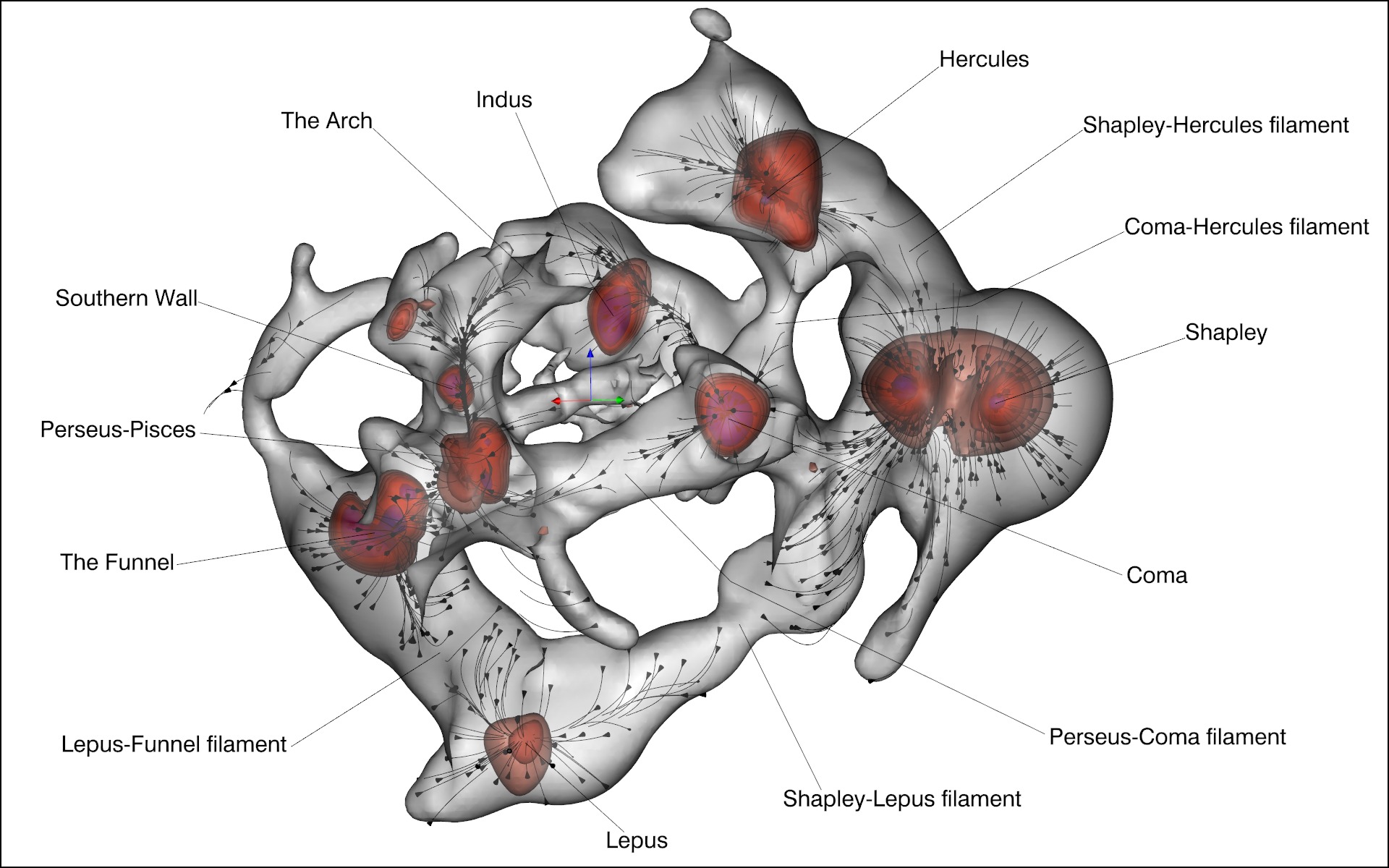}
\caption{The cosmic V-web from a rotated view.  Flow lines indicate local flows.  A similar image appears at 4:25 in the animated figure.}
\label{vweb_north}
\end{center}                                                                                                                                                                                                                                                                                                                                                                                                                                                                                                                                                                                        
\end{figure}

Figure~\ref{vweb_north} illustrates the same V-web structure from a rotated viewing position.  The connections between the Perseus-Pisces and Coma structures and between Hercules and Shapley are seen.  Also there is a clear view of our home position at the origin of the red, green, blue arrows.

It is to be appreciated that an individual galaxy only travels a few Mpc along the direction of a streamline in the age of the universe.  Peculiar velocities are only small perturbations on the cosmic expansion except in regions that have collapsed and virtualized.

\onecolumn                                                                                                                                                                                                                                                                                                                                                                                                                                                                                                                                                                                          
\twocolumn

\section{Cosmic V-Web}

The cosmic velocity web and the more familiar cosmic web defined by a redshift survey provide alternative descriptions of large scale structure.  Fundamentally they are in good agreement.  Redshift representations are straight forward at a basic level and can be rich in detail.  They suffer from boundary effects: in distance, with obscuration, and survey constraints.  Only certain kinds of galaxies are included and the linkage with dark matter is ambiguous.  With assumptions, velocity fields, the displacements of the observed galaxies from redshift locations can be estimated \citep{2006MNRAS.373...45E, 2008MNRAS.383.1292L, 2010MNRAS.406.1007L, 2016MNRAS.457..172L, 2012MNRAS.427L..35K, 2016MNRAS.457L.113K, 2013MNRAS.435.2065H}.  The Delaunay Tessellation Field Estimator has been used to infer velocity flows from a redshift survey \citep{2007MNRAS.382....2R}, demonstrating in particular the importance of radial outflows from voids.

The alternative description from the velocity field independent of information about the distribution of galaxies is complementary.  Each object with a measured peculiar velocity is a test particle responding to the gravity field at its position.  To date, only a modest fraction of galaxies drawn from a redshift survey contribute to a V-web analysis and the individual errors are large.  However, the long range correlation of individual velocities enables the inference of structures in regions of obscuration and on scales larger than probed with concurrent all-sky redshift surveys \citep{1987ApJ...313L..37D, 1999ApJ...522....1D, 1999ApJ...520..413Z, 2017NatAs...1E..36H}.  It is an important advantage that the density field is evaluated free of assumptions regarding the relationship with luminous objects.

\begin{figure}[!]
\begin{center}                                                                                                                                                                                                                                                                                                                                                                                                                                                                                                                                                                                      
\includegraphics[scale=.114]{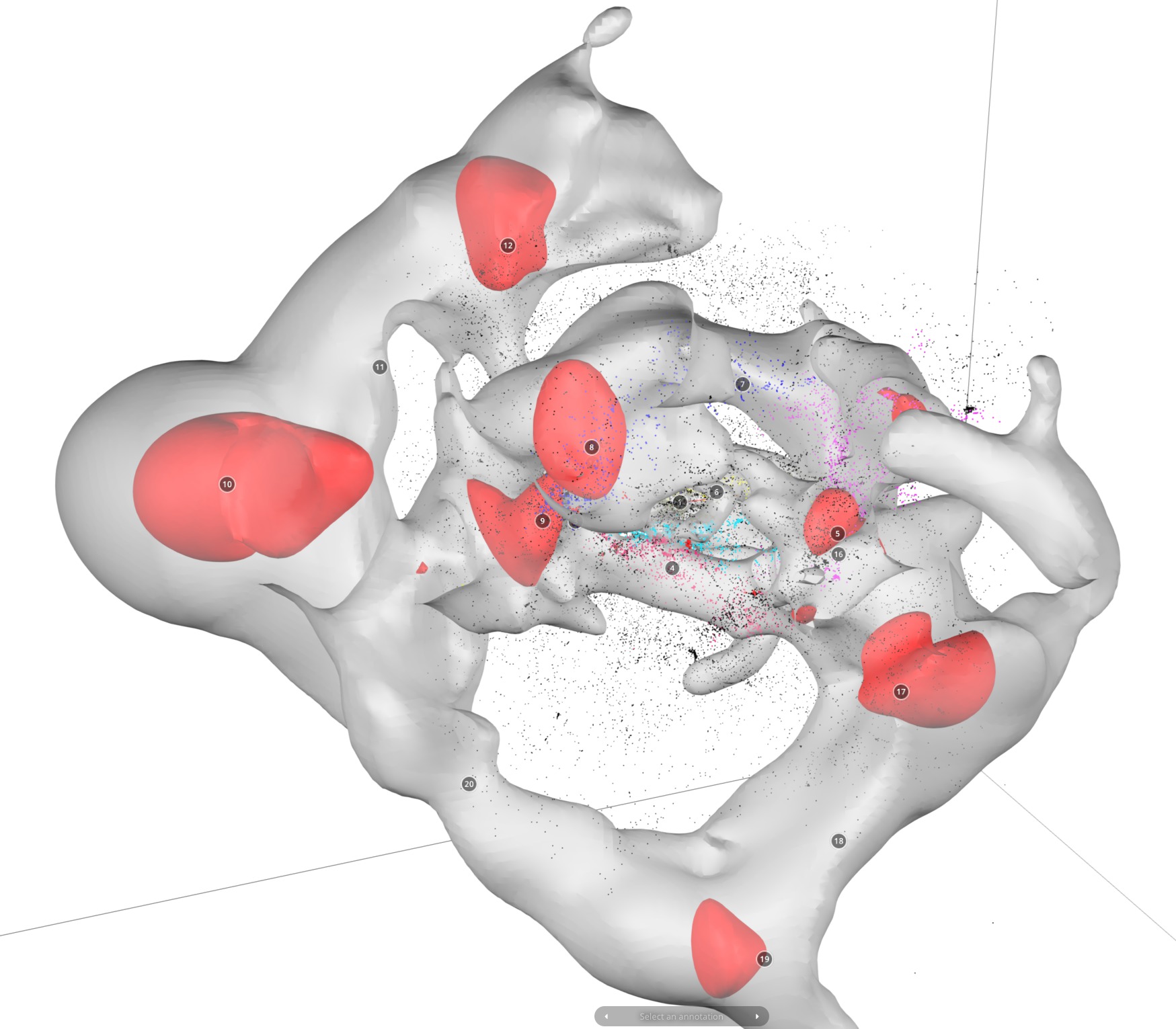}
\caption{The initial scene of the interactive figure of the {\it Cosmicflows-2} Wiener filter construction of the cosmic velocity web displayed in terms of filaments (grey surfaces) and knots (red surfaces).  Galaxy positions from the V8k catalog of redshifts are marked by spheroids of different colors: 1) red for a selection of clusters, 2) cyan, pink, and khaki for the three components of the Centaurus-Puppis-Perseus-Pisces filament, 3) yellow for the Centaurus-Virgo-Perseus-Pisces filament, 4) blue for the Centaurus-Arch-Perseus-Pisces filament, 5) magenta for Perseus-Pisces galaxies, 6) black otherwise.  The central signpost has 2,000~\kms-long arrows pointing to the SGX (red), SGY (green), SGZ (blue) directions.  The box is 64,000~\kms\ wide.  There are 20 annotated posts.  The model view focuses on a post upon selection.  Alternatively, the view will progress to each post in order by selecting the forward or backward arrows at the bottom of the image.  The model can also be mouse controlled with pan, zoom, rotate, and translate options.  Available at  https://skfb.ly/667Jr}
\label{skfb}
\end{center}                                                                                                                                                                                                                                                                                                                                                                                                                                                                                                                                                                                        
\end{figure}

The animated figure highlights the major features of the V-web as defined by the {\it Cosmicflows-2} compilation of distances and interpreted velocity field.   In addition to the animated figure we provide an interactive figure\footnote{https:/skfb.ly/667Jr}, introduced by Figure~\ref{skfb}.  In each representation
we draw attention to filaments that bridge between major attraction knots, particularly several that pass through the zone of obscuration.  There is a notable one between Lepus and Shapley. 
\citet{2016arXiv161104615K} have discussed a structure in Vela that we posit is a major element in the linkage between Lepus and Shapley.  
Another prominent but obscured structure runs between Perseus$-$Pisces and Coma and can also be seen in the redshift survey reconstruction by \citet{2016MNRAS.455.3169L}.  Three other more local filaments that run from the Centaurus Cluster in the Great Attractor region to the Perseus$-$Pisces complex have received scant attention and are discussed next.  We distinguish them by three-part names that identify their end points and the route they take.  Filamentary links between major knots is a fundamental expectation of the cosmic web \citep{1996Natur.380..603B, 1996MNRAS.281...84V, 2008LNP...740..409V}.

There will be reference below to an interactive figure of the V-web with superimposed galaxies from a redshift catalog.  The catalog, called "V8k"\footnote{Available at http://edd.ifa.hawaii.edu}  is restricted to a box of $\pm8,000$~\kms\ on the cardinal supergalactic axes \citep{2013AJ....146...69C}.  Galaxies in clusters that are important for our discussion are colored red.  Galaxies within the Perseus-Pisces structure are colored magenta.  Other features that are given colors are described below. 

\subsection{Centaurus-Puppis-PP Filament}

A substantial filament originates at the Centaurus Cluster, runs to the Antlia Cluster, then passes through the Milky Way plane at the south supergalactic pole, and proceeds onward to the Persus-Pisces complex.
Pieces of our Centaurus-Puppis-PP filament have been discussed by several authors but disjointly because of the interruption by obscuration.  The Antlia-Hydra Cloud in the {\it Nearby Galaxies Atlas} \citep{1987nga..book.....T} $-$ called the Hydra Wall by \citet{1998lssu.conf.....F} $-$ passes through the zone of obscuration as the Puppis filament
\citep{1992A&A...266..150K, 2016AJ....151...52S}, and emerges south of the Milky Way as the Lepus Cloud in the {\it Nearby Galaxies Atlas}.  The term Antlia strand \citep{2013AJ....146...69C} was introduced to identify the feature as one of five strands (filaments) emanating from the Centaurus Cluster.   

\begin{figure}[]
\begin{center}                                                                                                                                                                                                                                                                                                                                                                                                                                                                                                                                                                                      
\includegraphics[scale=.24]{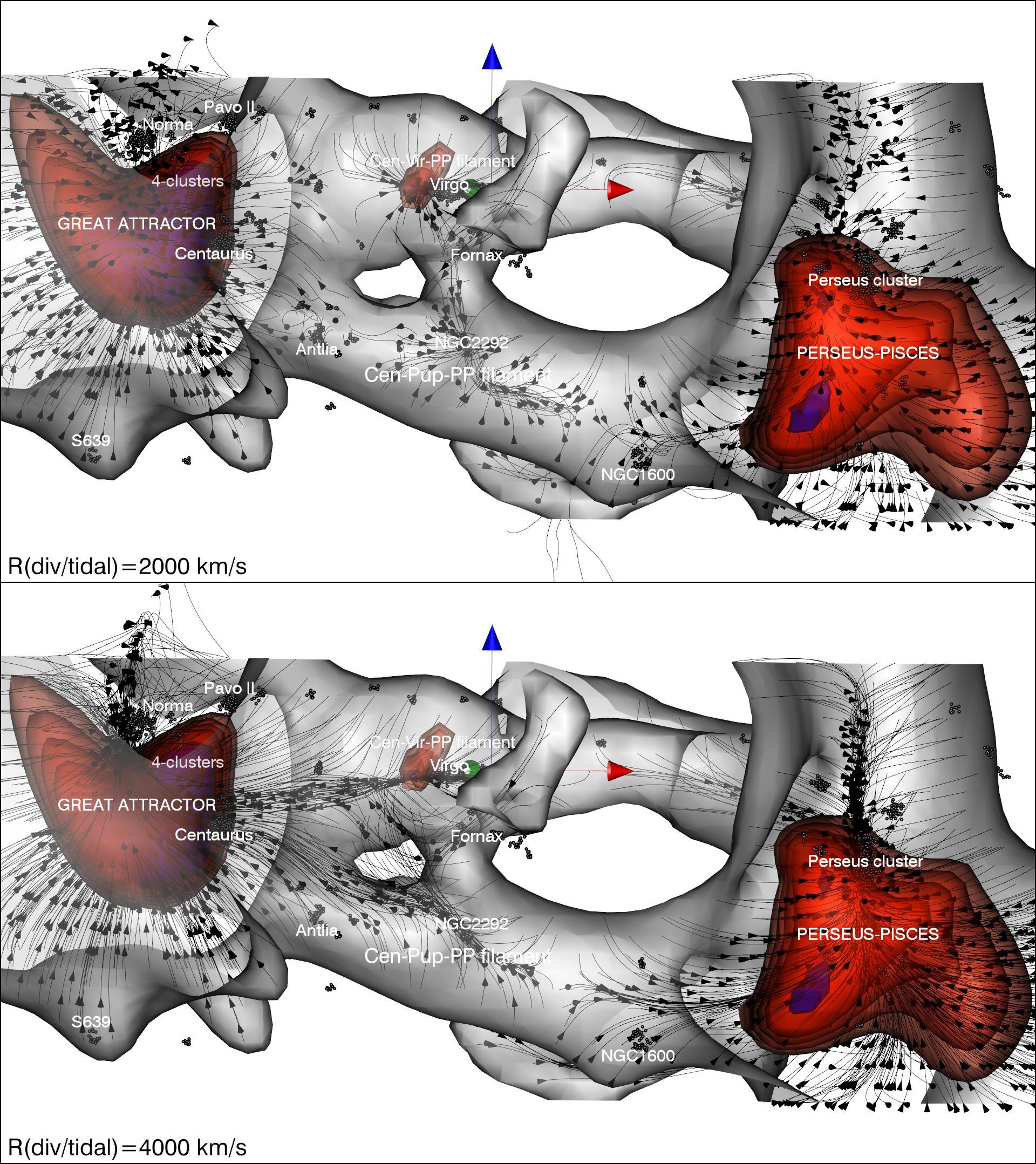}
\caption{The Centaurus-PP filament with two local/tidal cuts.  There is a representation in each panel of V-web surfaces of knots (red) and filaments (grey).  Local flow patterns derived as described in section~\ref{sec:velocity} are indicated by the flow lines with arrows.  In the top panel the local (divergent) velocities are calculated within spheres of radius 2,000 \kms\ while in the bottom panel the spheres are expanded to 4,000 \kms.  Flows are evident toward regional mass concentrations like the Virgo Cluster and the NGC 2292 Group in the top panel while there is greater organization in flows toward the major attractors in the bottom panel.  Important clusters within the observed slice are identified by name.  A variant of the bottom panel appears at 6:11 in the animated figure.}
\label{antlia}
\end{center}                                                                                                                                                                                                                                                                                                                                                                                                                                                                                                                                                                                        
\end{figure}

Figure~\ref{antlia} shows the V-web representation of the filament in question; it is the lower coherent structure between the Great Attractor and Perseus-Pisces in each of the two panels.  A bifurcation in flows is seen within the V-web filament roughly midway between the NGC 1600 and NGC 2292 groups where the abundance of observed galaxies becomes sparse and local flows split in direction.  Effectively, the boundary between the Laniakea \citep{2014Natur.513...71T} and Perseus$-$Pisces \citep{1988lsmu.book...31H} superclusters is being crossed.  Inspection of the two panels reveals variations in local flow patterns as a function of scale.  Local flows responding to density perturbations on small scales are seen to feed toward knots which serve as local attractors.  As the scale of influences is allowed to expand then coherence emerges in flows toward the major basins of attraction.

Giving consideration to the three-dimensional distribution of galaxies from the dense V8k redshift survey it appears that the Centaurus-Puppis-PP filament is part of a wall at the front boundary of the Sculptor Void  \citep{1998lssu.conf.....F}.  The wall is broken into three main filamentary strands, each extending all the way from the Centaurus Cluster to the vicinity of the Perseus-Pisces structure.  In \citet{2013AJ....146...69C} the initial parts leaving the Centaurus Cluster were referred to as the Antlia strand, already discussed, and the Southern Supercluster \citep{1956VA......2.1584D} strand with forks "$a$" and "$b$".  These separate features can be followed on the accompanying interactive figure.  In that model, galaxies in the main Centaurus-Puppis-PP filament are colored pink while the Southern Supercluster $a$ and $b$ strands are colored cyan and grey-green respectively.  In the V-web, only the Centaurus-Puppis-PP structure (emanating from the Antlia strand) has the eigenvalue properties of a filament given the threshold values specified in Section~\ref{sec:velocity}.  The other two structures (the two forks of the Southern Supercluster strand) are identified as sheets, expanding on two axes with compression normal to the void.  \citet{2013MNRAS.429.1286C} have shown that the shear field traces dominant filaments and nodes but is not sensitive to whisky features,

\subsection{Centaurus-Arch-PP Filament}

\begin{figure}[]
\begin{center}                                                                                                                                                                                                                                                                                                                                                                                                                                                                                                                                                                                      
\includegraphics[scale=.24]{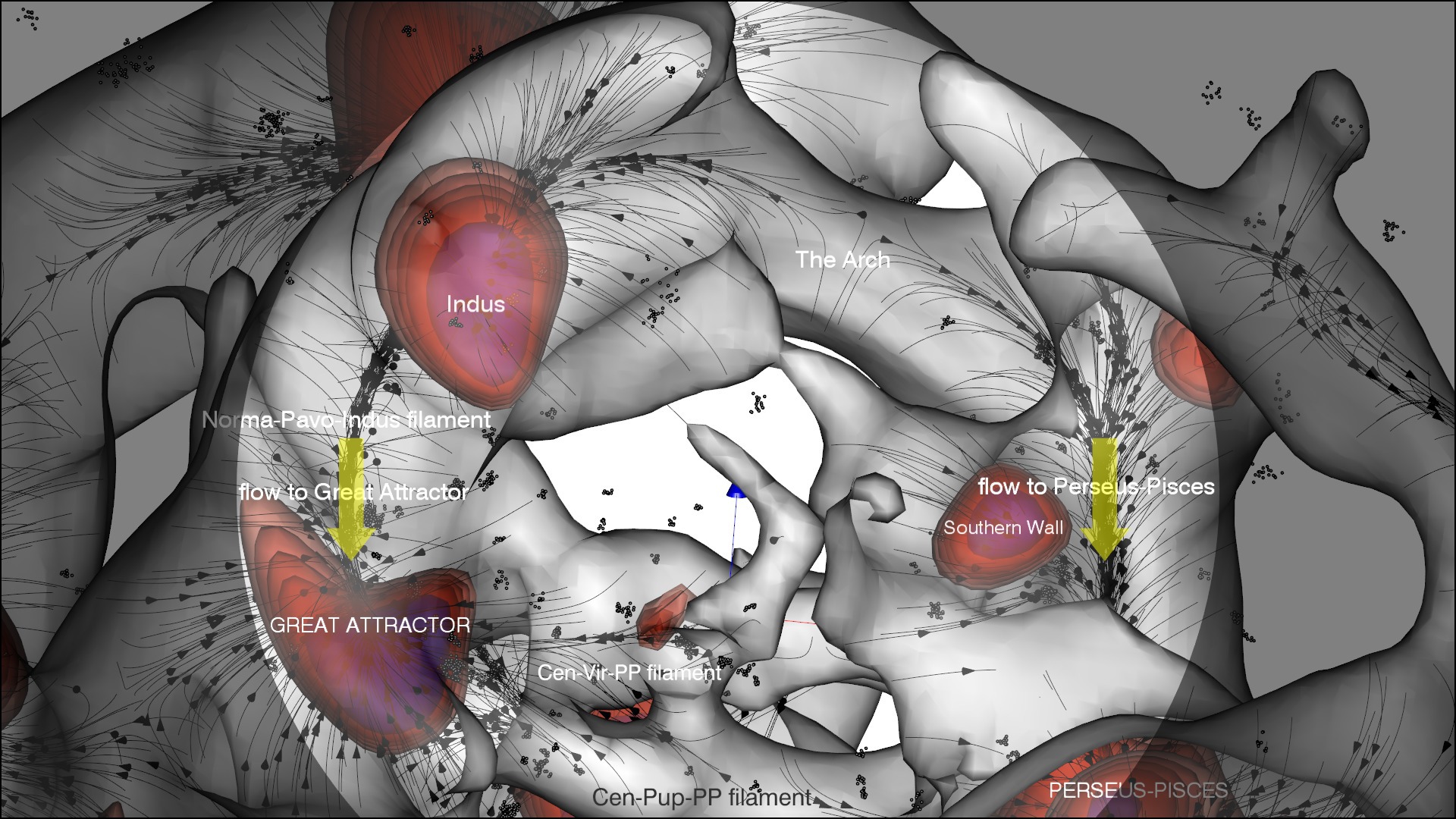}
\caption{A great filament runs from the Great Attractor at the heart of Laniakea Supercluster, through Norma-Pavo-Indus, across the Arch that forms a far boundary of the Local Void, to the Perseus-Pisces structure.  Midway along the Arch there is a point of bifurcation between local flows that head to Perseus-Pisces and local flows directed toward the Great Attractor.  This scene appears at 5:55 in the animated figure.}
\label{arch}
\end{center}                                                                                                                                                                                                                                                                                                                                                                                                                                                                                                                                                                                        
\end{figure}

A robust feature of the V-web analysis is an entity that we call the Arch.  This structure is an extension of the major Norma-Pavo-Indus filament ("NPI" strand in \citet{2013AJ....146...69C}) that originates near the Great Attractor convergence at the Centaurus Cluster and caps the Local Void in the supergalactic north before reaching the Perseus-Pisces region.  A view of it is shown in Figure~\ref{arch}.   This same view is seen at 5m 55s in the animated figure.  This scene illustrates the split in directions of local flows toward alternatively Perseus-Pisces and the Great Attractor via the Norma-Pavo-Indus filament.  Individual galaxies in this structure are represented in blue in the accompanying interactive figure. 

It has required the V-web analysis to clarify the importance of this connection between the Great Attractor and Perseus-Pisces regions.  The filament passes through galactic obscuration between the Centaurus and Norma clusters and then, after the easily followed Pavo-Indus thread, it folds back at its greatest distance of $\sim6,000$~\kms\ toward low galactic latitudes before reaching the Perseus-Pisces region in the vicinity of the NGC 7242 Cluster.

\subsection{Centaurus-Virgo-PP Filament}

\begin{figure}[]
\begin{center}                                                                                                                                                                                                                                                                                                                                                                                                                                                                                                                                                                                      
\includegraphics[scale=.25]{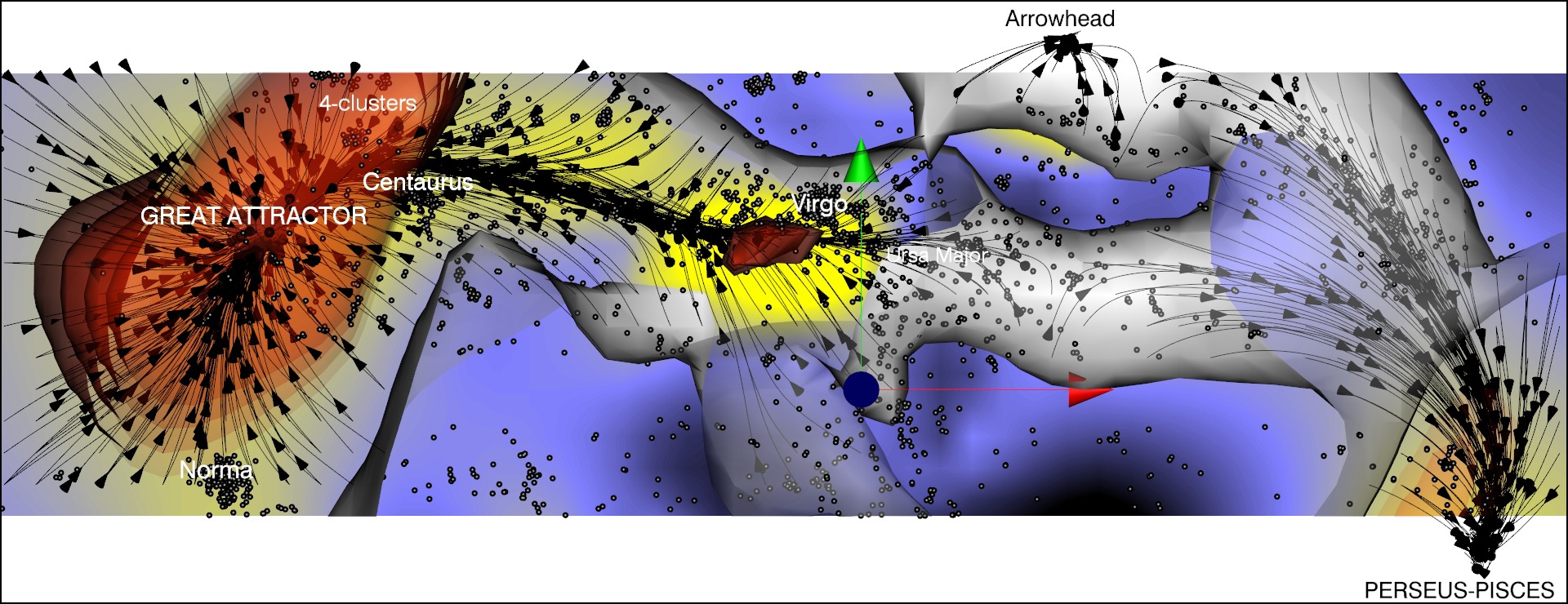}
\caption{Our relationship to the Great Attractor is shown.  Flow streamlines in our vicinity at the origin of the arrows pass by the Virgo Cluster and along the Virgo strand to arrive at the Great Attractor in the region of the Centaurus and Norma clusters.  A V-web filament iso-surface extends from the Great Attractor through Virgo to finally reach the Perseus-Pisces complex.  Individual galaxies in the selected volume are drawn from 2MRS.  Galactic obscuration lies in a horizontal belt coincident with the red arrow (supergalactic SGX, with SGZ normal to the figure) accounting for the lack of galaxies in that band.  Tones of blue through yellow and red map regions of increasing trace of the shear tensor that prescribes the linear density field in the Wiener filter model.  A variant of this scene appears at 7:09 in the animated figure.}
\label{virgo}
\end{center}                                                                                                                                                                                                                                                                                                                                                                                                                                                                                                                                                                                        
\end{figure}

It becomes apparent that the structure historically called the Virgo or Local Supercluster \citep{1953AJ.....58...30D} is only an appendage of larger structure.  In \citet{2013AJ....146...69C} the "Virgo strand" was another of five filaments emanating from the Centaurus Cluster (three of the others have been mentioned including the Norma-Pavo-Indus strand feeding into the Arch, the Antlia strand feeding the Centaurus-Puppis-PP filament, and the forked Southern Supercluster strand.)  In Figure~\ref{virgo} we note that this structure through Virgo can be followed all the way to the Perseus-Pisces region.  In the interactive figure galaxies in this filament are given the color yellow.   Toward the end of the path nearing the Perseus Cluster the individual galaxy sample is depleted by obscuration but the filamentary structure is robustly defined by the V-web.

Somewhat to the right of the feature labeled Ursa Major in Fig.~\ref{virgo} the local flows are seen to flip from toward the Great Attractor to toward Perseus-Pisces.
As Perseus-Pisces is approached the filament from Virgo converges with a filament from the Arrowhead mini-supercluster \citep{2015ApJ...812...17P}.

The convergence of local flow velocities at what we are calling the Great Attractor is seen in the figure to be between the Centaurus and Norma clusters.  The clusters Antlia, Hydra, Abell 3537, 3565, 3574, and S753 are also close by.  Linear theory fails to describe the details of this complex region.

\section{Discussion}

The description of large scale structure derived from the shear of the velocity field, the cosmic V-web, is blunt but informative. It is blunt in the sense that it does not access the fine details seen in the filagreed visual representations of the distribution of galaxies in redshift space, or the even more detailed renderings of simulations with large numbers of particles.  However, where the velocity information is of high enough quality, the V-web captures the essential elements of structure.  It captures the cores of attraction and repulsion in the knots and depths of voids.  With the separation into filaments and sheets the V-web identifies the dominant streams, whereas the relative importance of thread-like features are often not so obvious with redshift maps. 

The V-web (as presently derived) is blunt also because it is based on linear theory.  In reality, it can be expected that filaments and sheets are thinner than the linear theory representations.  Moreover, our comparison with redshift surveys (either 2MRS or V8k) do not account for the displacements of galaxies in redshift space with respect to physical space \citep{2016MNRAS.457L.113K, 2016MNRAS.455.3169L}.  Obvious consequences are visible on close inspection of the figures and the interactive figure.  The displacement of the V-web knot from the galaxies of the Virgo Cluster of $\sim400$~\kms\ is a particular example.  

Our reconstruction of knots, filaments, sheets and voids from peculiar velocity information alone should be compared in detail with constructions based on redshift surveys \citep{2007MNRAS.382....2R, 2016MNRAS.455.3169L, 2016MNRAS.457L.113K}.  We expect to make such comparisons but with a V-web construction based on the more extensive {\it Cosmicflows-3} compendium of distances.  The comparisons should also benefit from improved spatial resolution in the V-web analysis following from refinements that probe the mildly non-linear regime.

\begin{figure}[]
\begin{center}                                                                                                                                                                                                                                                                                                                                                                                                                                                                                                                                                                                      
\includegraphics[scale=.62]{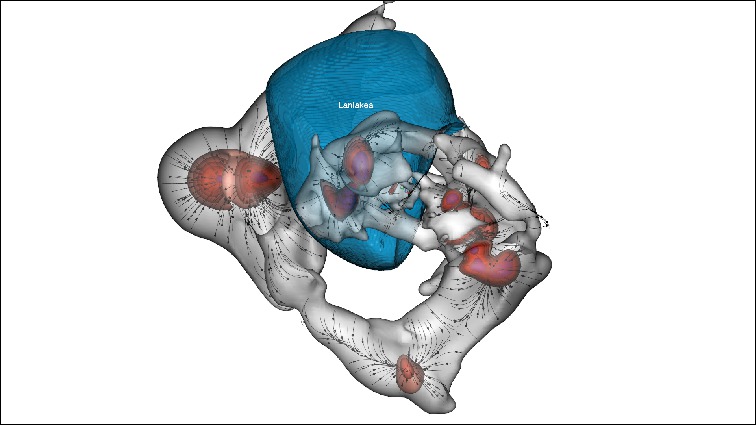}
\caption{The surface of the Laniakea Supercluster in blue is superimposed on the V-web.  Local flows within this surface terminate at the Great Attractor.  Local flows beyond this boundary are directed toward other basins of attraction.  See this scene at 7:30 in the animated figure.}
\label{laniakea}
\end{center}                                                                                                                                                                                                                                                                                                                                                                                                                                                                                                                                                                                        
\end{figure}

We have seen that V-web filaments can link between basins of attraction on various scales as anticipated by simulations \citep{1996Natur.380..603B, 2005MNRAS.359..272C}.  The Laniakea Supercluster is a basin of attraction on a scale of $10^{17} \Msun$ \citep{2014Natur.513...71T}.  Local flows on this scale and within the envelope of Laniakea gather in the Great Attractor region \citep{1988ApJ...326...19L}, near the nexus of the Norma and Centaurus clusters.  In Figure~\ref{laniakea} we see the domain of Laniakea Supercluster superimposed on the V-web based on velocities from {\it Cosmicflows-2} and our Wiener filter model.

\section{Conclusions}

The shear in the velocity field of galaxy flows provides a quantitative description of the knot, filament, sheet, void morphology of large scale structure.  The ensuing maps are necessarily coarse given the current completeness and accuracy of velocity information and the biases that arise with linear theory and the velocity displacements with redshift surveys.  The present construction rapidly deteriorates beyond $\sim 7,000$~\kms\ due to the paucity of data and large distance uncertainties.  However the major features of local large scale structure are recovered.  

In the animated figure and interactive figure several filaments connecting major centers of attraction are easily followed around the periphery of the observed volume.  There is a particularly prominent structure running from Perseus-Pisces to Lepus to Shapley.   Another runs from Perseus-Pisces to Coma.  Both of these filaments are followed as they pass through the zone of obscuration.  Then another filament of note runs from Hercules to Shapley.

Boring into the model where the data density and distance quality are highest, we find three significant filaments linking the Centaurus Cluster in the Great Attractor region to the Perseus-Pisces complex.   None of these are familiar in their entirety in the literature, primarily because all three pass through the obscuration of the Milky Way plane.  We submit that there is much to learn from the V-web.

\bigskip
We are indebted to the referee for many comments that have strengthened this paper.
Visualizations in this work are rendered with SDvision (Saclay Data Visualization) software.  Support has been provided over the years by the Institut Universitaire de France, the Israel Science Foundation (1013/12), the US National Science Foundation, the Space Telescope Science Institute, NASA, and the Jet Propulsion Lab.

\bigskip\noindent
Animated figure (video):

 http://vimeo.com/pomarede/vweb

\noindent
Interactive figure:

 https://skfb.ly/667Jr

\onecolumn                                                                                                                                                                                                                                                                                                                                                                                                                                                                                                                                                                                          
\twocolumn

\onecolumn                                                                                                                                                                                                                                                                                                                                                                                                                                                                                                                                                                                          
\twocolumn    

\onecolumn                                                                                                                                                                                                                                                                                                                                                                                                                                                                                                                                                                                          
\twocolumn

\onecolumn                                                                                                                                                                                                                                                                                                                                                                                                                                                                                                                                                                                          
\twocolumn    

\bibliography{paper}
\bibliographystyle{apj}

\end{document}